\documentclass{ws-procs9x6-cpt16}
\begin{document}

\newcommand{\refeq}[1]{(\ref{#1})}
\def\etal {{\it et al.}}

\def\aring{{\mathring{a}_{\mathrm{of}}^{(3)}}}
\def\twonu{{2\nu\beta\beta}}
\def\zeronu{{0\nu\beta\beta}}

\title{First Search for Lorentz and CPT Violation in\\
Double Beta Decay with EXO-200}

\author{L.J.\ Kaufman}

\address{Physics Department, Indiana University and CEEM\\
  Bloomington, IN 47405, USA}

\author{On behalf of the EXO-200 Collaboration}

\begin{abstract}
  This proceedings contribution reports the first experimental search for Lorentz- and CPT-violating signals specifically studying the effect of the Standard-Model Extension (SME) oscillation-free momentum-independent neutrino coupling operator in the double beta decay process.  The search has been performed using an exposure of 100 kg~yr of $^{136}$Xe with the EXO-200 detector.  No significant evidence of the spectral modification due to isotropic Lorentz violation was found.  A two-sided limit of $-2.65 \times 10^{-5}$ GeV $<$ $\aring$ $< 7.60 \times 10^{-6}$ GeV (90\% C.L.) is placed on the relevant coefficient within the SME. 
\end{abstract}

\bodymatter

\section{Introduction}
The Standard Model of particle physics 
assumes complete invariance under Lorentz transformations which leads to invariance under CPT transformation.  The observation of the violation of either of these symmetries would imply the observation of new physics beyond the Standard Model.

The Standard-Model Extension (SME), developed by Kosteleck\'y \etal,\cite{exo10,exo11,exo12} provides a framework to test for violations of these symmetries with current experimental technologies.  
The operators that couple to neutrinos in the SME affect the flavor oscillation properties, neutrino velocity, or phase space.\cite{exo16,exo17}  
There exists an operator in the SME that couples to neutrinos that is momentum independent and does not affect neutrino oscillation (oscillation free) and is unobservable to long-baseline neutrino experiments.\cite{exo8}  This renormalizable Lorentz-violating operator, known as the {\it countershaded} operator, has mass dimension three and also breaks CPT.  The corresponding countershaded coefficient has four components, one time-like $(\aring)_{00}$ and three space-like $(\aring)_{1m}$, with $m = 0, \pm1$.  A nonzero value of $(\aring)_{00}$ would produce small deviations in the shape of the energy spectrum for single or double beta decay and can be searched for experimentally.  
In this work we employ a new method to explore for the first time the effects of the countershaded coefficient $(\aring)_{00}$ on a wide energy range of the double beta decay spectrum.

\section{Detector description}
The EXO-200 detector was built to measure the two-neutrino double beta decay ($\twonu$) spectrum of $^{136}$Xe and to search for the neutrinoless version ($\zeronu$) by measuring the electron energy sum spectrum from these processes with high precision.  EXO-200 is a good candidate detector to search for the effects of the time-like component on double beta decay due to the low background of the experiment and ability to measure precisely the spectral shape.

The EXO-200 detector is described in detail elsewhere.\cite{exo30}  In summary, the detector is made up of two back-to-back cylindrical time projection chambers that share a central cathode.  
The detector is filled with $\sim$ 175 kg liquid xenon (LXe) that has been enriched to 80.6\% $^{136}$Xe.
Ionizing particle interactions in the LXe produce both scintillation light and ionization electrons.  Both signals are collected in the detector to determine the energy and the 3D position of the event cluster.
The detector system is located in a clean room underground 
at the Waste Isolation Pilot Plant near Carlsbad, NM, USA.

\begin{figure}[b]
\begin{center}
\includegraphics[width=\hsize]{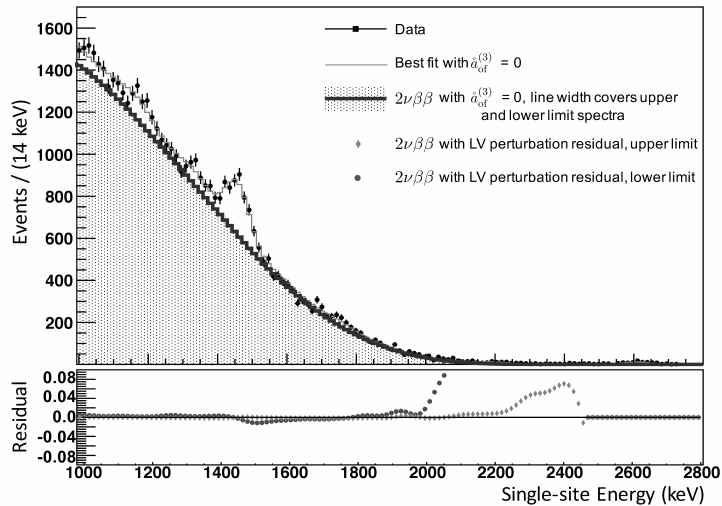}
\end{center}
\caption{The SS data for the case of zero perturbation due to Lorentz violation are shown with the best overall fit in the energy spectrum, with the best fit of the perturbed $\twonu$ spectra at the upper and lower 90\% CL bounds of the fit for Lorentz violation indicated by the width of the line.  
  The fractional residual difference in the total number of counts per bin between the $\twonu$ spectrum in the case of no Lorentz violation and the upper and lower bound cases is shown on the lower figure.}
\label{Spectrum}
\end{figure}

\section{Analysis method and search strategy}
The coupling of a neutrino to the countershaded operator alters the neutrino momentum 
which modifies the double beta decay transition amplitude as well as the neutrino dispersion relation.\cite{exo33}  This leads to a differential decay rate in terms of the kinetic energies of the two emitted electrons.  
The coefficient $\aring$ is the parameter of interest 
and is related to the time-like component of the countershaded operator coefficient by $\aring = (a^{(3)}_{\mathrm{of}})_{00}/\sqrt{4\pi}$.

The decay rate can be separated into two distinct parts; one corresponding to the standard $\twonu$ process,\cite{exo35} and a second term corresponding to the perturbation of the $\twonu$ spectrum due to the coupling of neutrinos to the Lorentz-violating operator (LV$\beta\beta$).  Precision measurements of the $\twonu$ spectrum require $|\aring| \ll 1$,\cite{exo36} so the total decay rate can be expressed as an addition of the two separate rates through a perturbation.

The SME coefficient $\aring$ only affects the phase-space factor perturbation in the decay rate. 
This can be related to an effective decay rate of the Lorentz-violating perturbation to the $\twonu$ spectrum.  
More details on the analysis method can be found in Ref.~\refcite{exoLV}.

This analysis uses the same event reconstruction and fitting techniques as described in detail in previous publications.\cite{exo36,exo37,exo38}  The same dataset is also used, consisting of a total exposure of 100 kg~yr.  
Probability density functions (PDFs) for the $\twonu$ and LV$\beta\beta$ signals and expected backgrounds are produced using the Geant4-based\cite{exo39} EXO-200 simulation software, which is described in detail elsewhere.\cite{exo36}  Both data and PDFs are separated into single-site (SS) and multi-site (MS) spectra according to the number of separate charge clusters observed.  A simultaneous fit to the SS and MS spectra is performed to constrain both the $\beta$-like signal events, which are primarily SS, and the $\gamma$-like backgrounds, which are primarily MS.

The analysis region for this search is between 980 and 9800 keV. PDFs for expected backgrounds and the $\twonu$ and LV$\beta\beta$ signal functions are fit to the selected data by minimizing the negative-log likelihood function.  A profile likelihood scan over the number of LV$\beta\beta$ integral counts added to or subtracted from the standard $\twonu$ spectrum is used to obtain limits at the 90\% confidence level (CL).

Several studies were performed on the background model to obtain gaussian constraints on systematic uncertainties for the negative-log likelihood fit.  
More information about the constraints is given in 
Refs.\ \refcite{exoLV} and \refcite{exo38}.

\section{Results}
A profile likelihood scan was performed over both positive and negative contributions of LV$\beta\beta$ counts, altering the standard $\twonu$ with both positive and negative values of $\aring$. 
The number of LV$\beta\beta$ counts at the 90\%~CL was converted into limits on the parameter of interest of $-2.65 \times 10^{-5}$ GeV $<$ $\aring$ $< 7.60 \times 10^{-6}$ GeV.  The perturbed $\twonu$ spectra with $\aring$ at the 90\% CL limits are shown in Fig.~\ref{Spectrum}.

In conclusion, we report on the first experimental search in double beta decay for the isotropic component of the coefficient describing the momentum-independent and oscillation-free operator coupling to neutrinos in the SME.  We detect no significant signal from studying the potential shape deviation from the standard $\twonu$ spectrum and set limits on the magnitude of this coefficient.

\end{document}